\documentclass[aps,pra,groupedaddress,twocolumn,showpacs]{revtex4}
\usepackage{graphicx,psfrag,amsmath,calc,amssymb}

\begin{document}

\title{BCS to BEC Quantum Phase Transition in Spin Polarized Fermionic Gases}
\author{S. S. Botelho}
\author{C. A. R. S\'{a} de Melo}
\affiliation{School of Physics, Georgia Institute of Technology,
             Atlanta Georgia 30332}

\date{\today}

\begin{abstract}
We discuss the possibility of a quantum phase transition in 
ultracold spin polarized fermionic gases which exhibit a 
$p$-wave Feshbach resonace. We show that when fermionic atoms 
form a condensate that can be externally
tuned between the BCS and BEC limits, the zero temperature 
compressibility and the spin susceptibility of the fermionic gas 
are non-analytic functions of the two-body bound state energy.
This non-analyticity is due to a massive rearrangement of
the momentum distribution in the ground state of the 
system. Furthermore, we show that the low temperature 
superfluid density is also non-analytic, and exibits a dramatic 
change in behavior when the critical value of the bound 
state energy is crossed.

\end{abstract}

\maketitle

%%%%%%%%%%%%%% Introduction %%%%%%%%%%%%%%%

{\it Introduction:}
Recent experiments in cold fermionic gases have shown that 
$s$-wave magnetic field induced Feshbach resonances can be 
used to form diatomic molecules of 
$^{40}{\rm K}$ \cite{jin-04} and 
$^6{\rm Li}$ \cite{zwierlein-03, bourdel-04, kinast-04}, which 
undergoe Bose-Einstein condensation (BEC) on the higher magnetic field
side of the resonance. On the lower magnetic field side of the
resonance, it has also been established that Cooper pairing
takes place and a BCS condensate is formed.
These studies in cold fermionic gases led to the first 
experimental realization of the theoretically proposed BCS-to-BEC crossover
in three dimensional continuum $s$-wave 
superfluids \cite{sademelo-93, engelbrecht-97}.
Three early theoretical works that considered the possibility of
$s$-wave superfluidity in the context of (what is known today as) the
BCS-to-BEC crossover should be highlighted. The first is   
by Eagles \cite{eagles-69}, where the  
possibility of pairing without condensation is described 
in a continuum model in the context 
of superconductors with low carrier concentration.
The second is Leggett's seminal work \cite{leggett-80a}, 
in which the $T = 0$ $s$ and $p$-wave BCS-to-BEC evolution are 
discussed as a crossover phenomenon in the context of a 
variational ground state wavefunction. 
And the third is the work of Nozieres-Schmitt-Rink \cite{nozieres-85},
where the $s$-wave BCS-to-BEC crossover in a lattice  
is described. Furthermore, much of the 
theoretical \cite{holland-01, griffin-02} and 
experimental \cite{jin-04, zwierlein-03, bourdel-04, kinast-04}
efforts that followed described only the BCS-to-BEC crossover in 
$s$-wave systems.

In this manuscript, we present a functional
integral analysis of the BCS-to-BEC evolution in $p$-wave fully 
spin-polarized Fermi gases, where $p$-wave Feshbach resonances have 
already been observed \cite{jin-03, zhang-04}. We show that a quantum 
phase transition takes place when the chemical potential crosses a 
critical value, instead of the usual smooth BCS-to-BEC crossover 
that occurs in $s$-wave superfluids~\cite{duncan-00}.  
The atomic compressibility and the spin-susceptibility of the Fermi gas 
are computed and are shown to be non-analytic in the $p$-wave case, 
as a consequence of a major rearrangement in the momentum distribution 
as the critical point is approached.
This non-analytic behavior suggests the occurrence of a quantum phase 
transition, which is further confirmed by a discontinuous 
change in the temperature dependence of the superfluid density of the 
gas at the transition point, which goes from power-law on the
BCS side of the resonance to exponential on the BEC side of the resonance.

We study the case of two-dimensional systems, which can be 
prepared experimentally through the formation of a one-dimensional 
optical lattice, where tunnelling between lattice sites is suppressed 
by a large trapping potential. The form of the trapping potential can be
chosen to be
$V_{{\rm trap}} = - I_0 \exp[-2 (x^2 + y^2)/w^2] \cos^2 (k_z z)$, 
where $2\pi / k_z$ is the wavelength of the light 
used in the laser beam.
We assume that the width $w$ is such that $w \gg \lambda_F$, where 
$\lambda_F = 2\pi/k_F$ 
is proportional to the interparticle spacing of a Fermi gas with Fermi 
wavevector ${\bf k}_F$, such that the problem is essentially two-dimensional.

%%%%%%%%%%%%%% Hamiltonian %%%%%%%%%%%%%%%

{\it Hamiltonian:} 
We study a two-dimensional continuum model of {\it spin polarized}
(all atoms in the same hyperfine state) fermionic atoms of mass $m$ 
and density $n = k_F^2 / 4\pi$. In the presence of an external magnetic field
${\bf h}$, the system is described by the Hamiltonian ($\hbar=k_B=1$)
\begin{equation}
{\cal H}=\sum_{\bf k}\xi_{\bf k}\psi_{{\bf k}\uparrow}^\dagger
\psi_{{\bf k}\uparrow} + {1 \over 2} \sum_{{\bf k},{\bf k}',{\bf q}} V_{{\bf k}{\bf k}'}
b_{{\bf k}{\bf q}}^\dagger b_{{\bf k}'{\bf q}} ,
\end{equation}
where $b_{{\bf k}{\bf q}}=\psi_{-{\bf k}+{\bf q}/2\uparrow}
\psi_{{\bf k}+{\bf q}/2\uparrow}$ 
and
$\xi_{\bf k}= \epsilon_{\bf k}  - \tilde\mu$, 
with
$\epsilon_{\bf k} = k^2/2m$, and  
$\tilde\mu = \mu + g_{{\tilde z} {\tilde z}} \mu_B h_{\tilde z}$.
The direction of the magnetic field ${\bf h}$, 
which was chosen to define the spin quantization axis ${\tilde z}$, 
need not to coincide with the spatial direction $z$ of propagation of the 
laser beam.

The interaction potential is approximated by the following separable 
function in ${\bf k}$-space,
\begin{equation}
V_{{\bf k}{\bf k}'} = -\lambda 
\Gamma ({\bf k}) \Gamma ({\bf k}') ,
\end{equation}
where $\lambda$ is the interaction strength and 
$
\Gamma ({\bf k}) = h(k) \cos(\varphi) ,
$
where the funcion 
$
h (k) = (k/k_1) / [1+k/k_0]^{3/2}
$
controls the range of the interaction, $\varphi$ is the 
momentum angle in polar coordinates, and  
$R_0 \sim k_0^{-1}$ plays the role of the interaction 
range. The functional form of $\Gamma ({\bf k})$ 
can be shown to produce the correct asymptotic behavior 
at small and large momenta~\cite{duncan-00}, and its
angular dependence reflects equal contributions from 
the angular momentum channels $\ell = \pm 1$.
In the limit of small momenta, this approach is identical 
to the $T$-matrix formalism \cite{leggett-80a}, 
but has the added advantage of making unnecessary to introduce 
the scattering length as a relevant parameter, 
which is quite problematic in two-dimensions~\cite{randeria-89}.
The BCS-BEC evolution can be safely analyzed 
provided that the system is dilute enough ($k_F^2 \ll k_0^2$),
i.e., the square of the
interparticle spacing ($\sim k_F^{-1}$) is much larger 
than the square of the interaction range ($\sim k_0^{-1}$).
Throughout the manuscript, we choose to scale all 
energies with respect to the Fermi energy 
$\epsilon_F = k_F^2 / 2m$ and all momenta with respect 
to $k_F$.

%%%%%%%%%%%%%% Effective Action %%%%%%%%%%%%%%%

{\it Effective Action:} 
The partition function $Z$ at a temperature $T = \beta^{-1}$ is written 
as an imaginary-time functional integral with action
$
S=\int_0^\beta d\tau[\sum_{\bf k} 
\psi_{{\bf k}\uparrow}^\dagger(\tau)
\partial_\tau \psi_{{\bf k}\uparrow}(\tau) + {\cal H}].
$
Introducing the usual Hubbard-Stratonovich field $\phi_{\bf q}(\tau)$, 
which couples to $\psi^\dagger\psi^\dagger$, and integrating out the fermionic 
degrees of freedom, we obtain 
\begin{equation}
Z=\int {\cal D}\phi{\cal D}\phi^* \,
\exp(-S_\mathrm{eff}[\phi,\phi^*]) ,
\end{equation}
with the effective action given by
$$
S_{\rm{eff}} = \int_0^\beta \!\! d\tau
\left[
U (\tau) + \sum_{{\bf k}, {\bf k}^\prime} 
\left({\xi_{\bf k}\over 2} \delta_{{\bf k}, {\bf k}^\prime} -
{\rm Tr} \ln {1\over 2}\mathbf{G}_{{\bf k},{\bf k}^\prime}^{-1}(\tau) \right)
\right],
$$
where 
$
U (\tau) = \sum_{\bf k}
|\phi_{\bf k}(\tau)|^2 / (2\lambda)
$
and $\mathbf{G}_{{\bf k}, {\bf k}^\prime}^{-1}(\tau)$ 
is the (inverse) Nambu matrix,
\begin{equation}
\mathbf{G}_{{\bf k},{\bf k'}}^{-1}(\tau) = 
\left(
\begin{array}{cc}
-(\partial_\tau + \xi_{\bf k})\delta_{{\bf k},{\bf k}^\prime} & 
\Lambda_{ {\bf k}, {\bf k}^\prime} (\tau) \\[3mm]
\Lambda^*_{ {\bf k}^\prime, {\bf k}} (\tau) &
-(\partial_\tau - \xi_{\bf k}) \delta_{{\bf k},{\bf k}^\prime} 
\end{array}
\right),
\end{equation}
with
$
\Lambda_{{\bf k},{\bf k}^\prime} (\tau) = 
\phi_{{\bf k} - {\bf k}^\prime} (\tau) 
\Gamma ( ({{\bf k} + {\bf k}^\prime})/2) .
$ 
%%
%

%%%%%%%%%%%%%% Saddle-point equation %%%%%%%%%%%%%%%

{\it Saddle Point Equation:} 
After Fourier transforming from imaginary time to Matsubara frequency 
$
(ik_n = i (2n+1) \pi/\beta)
$
and performing the frequency sum, the saddle point condition
$
[\delta S_{\rm eff} / 
\delta \phi^*_{\bf q}(\tau')]_{\Delta_0} = 0
$
can be cast in the form of the familiar order 
parameter equation,
\begin{equation}
{1 \over \lambda}  = 
\sum_{\bf k} {\Gamma^2({\bf k}) \over 2 E_{\bf k}}
\tanh \left( \beta E_{\bf k} \over 2 \right) ,
\end{equation}
where
$
E_{\bf k} = \sqrt{\xi_{\bf k}^2 + |\Delta_{\bf k}|^2}
$
is the quasiparticle excitation energy, and 
$\Delta_{\bf k} = \Delta_0 \Gamma ({\bf k})$ plays the role
of the order parameter function. 
We eliminate the interaction strength $\lambda$ in favor 
of the two-body bound state energy $E_b(h_{\tilde z})$ in vacuum 
(and in the presence of a magnetic field)
by using the relation
$
1/\lambda  = 
\sum_{\bf k} \Gamma^2({\bf k}) / 
(2\epsilon_{\bf k} - \tilde E_b),
$
where 
$
\tilde E_b = E_b(h_{\tilde z}) + 
2 g_{{\tilde z}{\tilde z}} \mu_B h_{\tilde z} .
$
The renormalized gap equation in terms of $\tilde E_b$ 
then takes the form
\begin{equation}
\sum_{\bf k} \Gamma^2({\bf k}) \left[
{1 \over 2 \epsilon_{\bf k} - \tilde E_b} - 
{\tanh(\beta E_{\bf k} / 2) \over 2 E_{\bf k}}
\right] = 0 .
\end{equation}
%%
%

%%%%%%%%%%%%%% Number equation %%%%%%%%%%%%%%%

{\it Number Equation:}
Using the relation 
$
N = -\partial\Omega / \partial\mu
$
and the saddle point approximation for the thermodynamic 
potential, 
$
\Omega_0 = S_{\rm eff}[\Delta_0] / \beta,
$
one can write the number equation as
$
N_0 = \sum_{\bf k} n_{\bf k} ,
$
where the momentum distribution $n_{\bf k}$ is given by
\begin{equation}
\label{eqn:nk}
n_{\bf k} = {1 \over 2} \left[
1 - {\xi_{\bf k}\over E_{\bf k}} 
\tanh \left( \beta E_{\bf k} \over 2 \right) \right].
\end{equation}
Thus, at $T=0$, the saddle point and number equations reduce to 
$
\sum_{\bf k} \Gamma^2({\bf k})
[(2\epsilon_{\bf k} - \tilde E_b)^{-1} - 
(2 E_{\bf k})^{-1}] =0 
$
and
$
N_0 = \sum_{\bf k} (1 - \xi_{\bf k} / E_{\bf k})/2,
$
respectively.
The solutions for $\Delta_0$ and $\tilde\mu$ at $T = 0$
as functions of the binding energy $\tilde E_b$ 
in the case of $p$-wave pairing symmetry 
are plotted in Fig. 1 for $k_0 = k_1 = 10 k_F$.
The point $\tilde\mu = 0$ is achieved for 
$\tilde E_b = -1.087\epsilon_F$ and corresponds to 
$\Delta_0 = 19.063\epsilon_F$.
\begin{figure}
\begin{center}
\includegraphics[width=6.5cm]{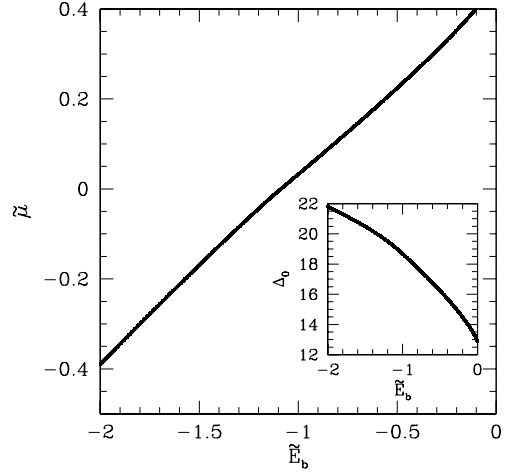}
\end{center}
\vspace{-5mm}
\caption{\small Universal plot (for any magnetic field $h_{\tilde z}$) of 
$\tilde\mu = \mu + g_{{\tilde z} {\tilde z}} \mu_B h_{\tilde z}$ 
and order parameter amplitude $\Delta_0$ (inset) as functions 
of $\tilde E_b = E_b(h_{\tilde z}) + 
2 g_{{\tilde z}{\tilde z}} \mu_B h_{\tilde z}$ 
for $k_0 = k_1 = 10 k_F$ in the spin polarized $p$-wave case. 
All quantities are in units of $\epsilon_F$.}
\end{figure}
%%
%

%%%%%%%%%%%%%% Gaussian Fluctuations %%%%%%%%%%%%%%%

\begin{figure*}
\begin{center}
\psfrag{kx}{$k_x$}
\psfrag{ky}{$k_y$}
\psfrag{nk}{$n_{\bf k}$}
\includegraphics[width=5.0cm]{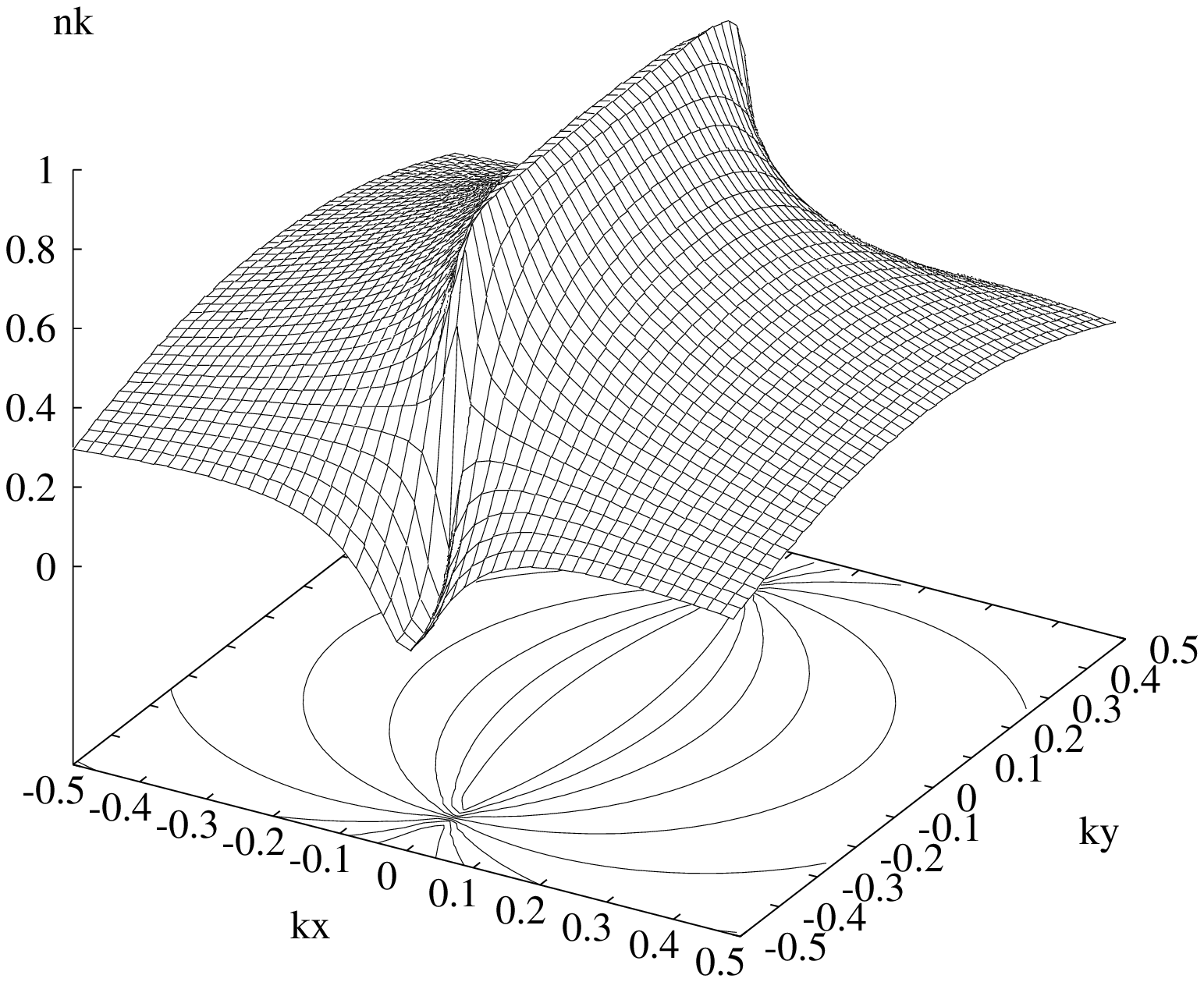}
\hspace{3mm}
\includegraphics[width=5.0cm]{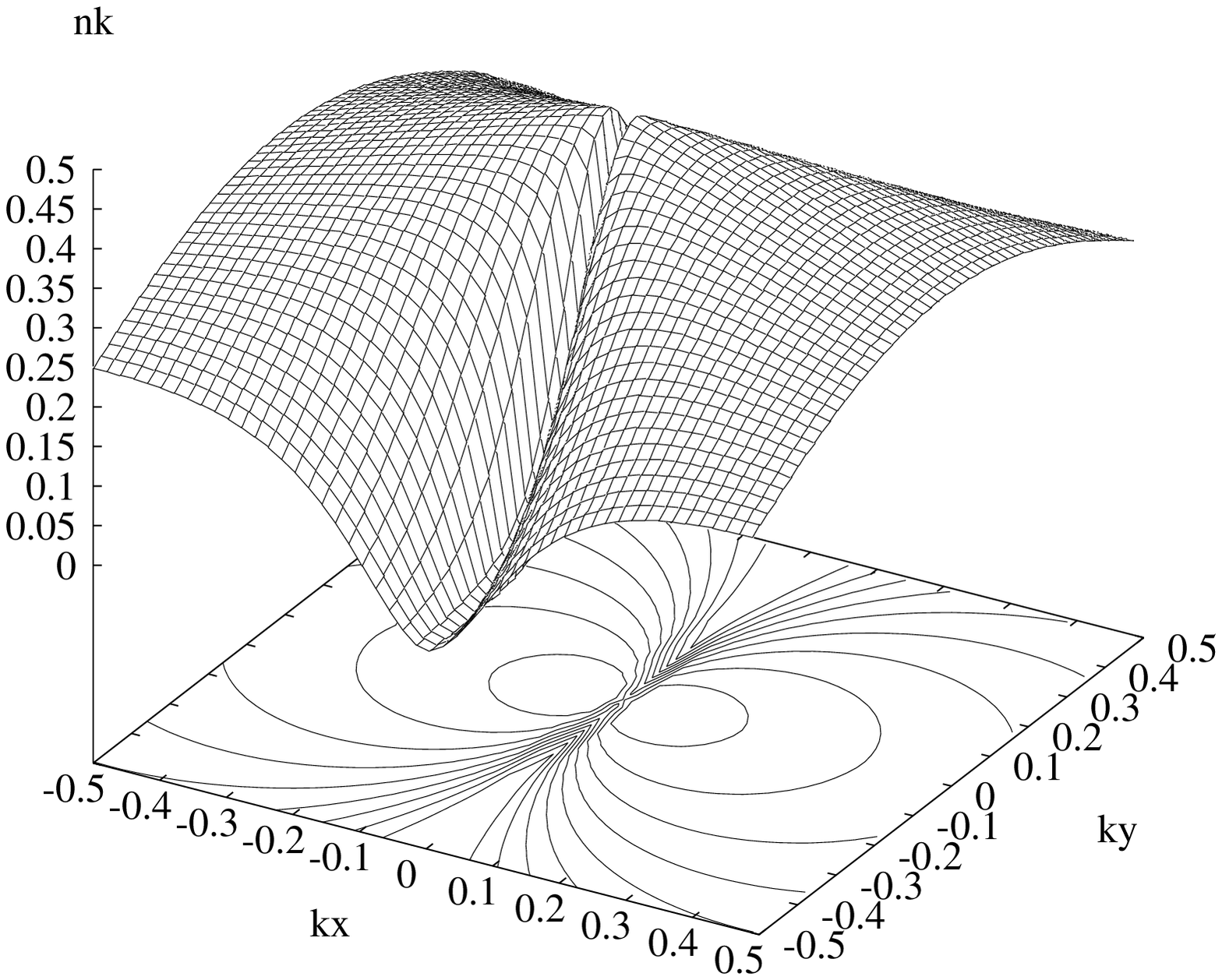}
\hspace{3mm}
\includegraphics[width=5.0cm]{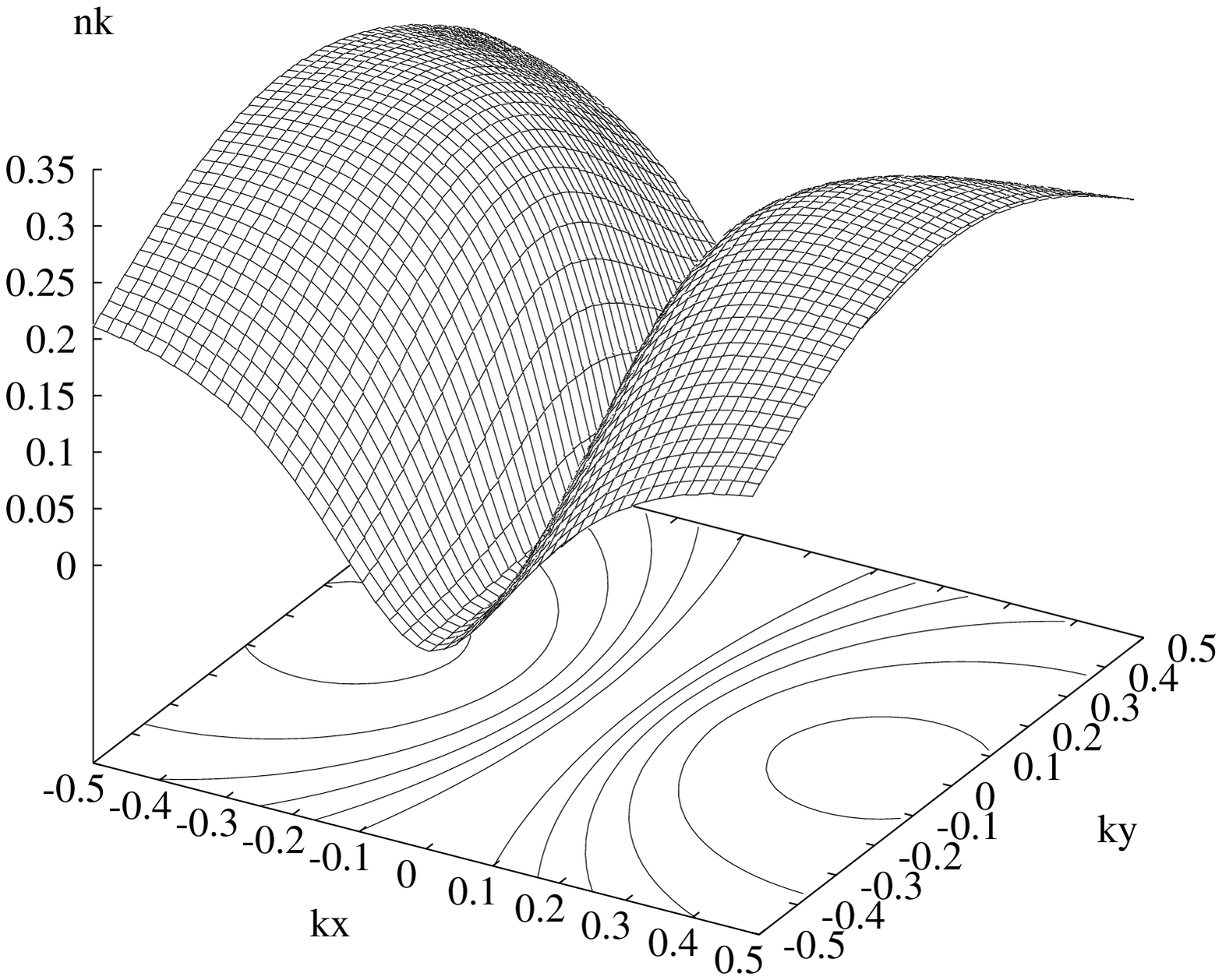}
\end{center}
\vspace{-5mm}
\caption{\small Plot of the momentum distribution $n_{\bf k}$ 
in the spin polarized $p$-wave case for (a) $\tilde\mu=0.15\epsilon_F$, 
(b) $\tilde\mu=0$ and (c) $\tilde\mu=-0.15\epsilon_F$. 
Notice the collapse of the two Dirac points 
when $\tilde\mu$ crosses zero.}
\end{figure*}

{\it Gaussian Fluctuations:}
We now investigate the effect of Gaussian fluctuations in the 
pairing field $\phi_{\bf q}(\tau)$ about the static saddle point 
value $\Delta_0$. Assuming 
$
\phi_{\bf q}(\tau) = \Delta_0 \delta_{{\bf q},0} + \eta_{\bf q}(\tau)
$
and performing an expansion in $S_{\rm eff}$ to quadratic order 
in $\eta$, one obtains
\begin{equation}
S_{\rm Gauss} = S_0[\Delta_0] + {1\over 2}
\sum_q \underline\eta^{\dagger}(q) 
{\bf M}(q) \underline\eta(q) ,
\end{equation}
where  $S_0$ is the saddle point action, the 
vector 
$
\underline\eta (q)
$
is such that
$
\underline\eta^{\dagger}(q) = [\eta^*(q) , \eta(-q)],
$
and $q \equiv ({\bf q}, iq_m)$, where 
$iq_m = i2m\pi/\beta$ is a bosonic Matsubara frequency.
The $ 2\times 2$ matrix ${\bf M}(q)$ is the inverse 
fluctuation propagator. 

The Gaussian fluctuation term in the effective action leads 
to a correction to the thermodynamic potential, which can 
be rewritten as 
$\Omega_{\rm Gauss} = \Omega_0 + \Omega_{\rm fluct}$,
with
$
\Omega_{\rm fluct} = 
\beta^{-1} \sum_q \ln \det [{\bf M}(q)].
$
Therefore, using the relation 
$
N=-\partial\Omega / \partial\mu,
$
one can write the corrected number equation as
$
N_{\rm Gauss} = N_0 + N_{\rm fluct},
$
where $N_0$ is the saddle-point level number of 
particles given above, and 
\begin{equation}
N_{\rm fluct} = - { \partial \Omega_{\rm fluct} \over  \partial \mu } =
T \sum_{\bf q}\sum_{i q_n}
\left[ { 
{-\partial ({\rm det} {\bf M})/ \partial \mu } \over 
{\rm det} {\bf M} ({\bf q}, iq_n)
}
\right] .
\end{equation}
At low $T$, the Goldstone mode $\omega = c |{\bf q}|$
dominates the contribution to $N_{\rm fluct}$, leading to
\begin{equation}
N_{\rm fluct} \sim  
- {L^2 \over 2\pi} \zeta(3)
{1 \over c^3}  
{\partial c\over \partial \mu}
T^3 ,
\end{equation}
which vanishes in the limit of $T \to 0$. Therefore, 
analogously to the three-dimensional $s$-wave case~\cite{engelbrecht-97}, 
Eq.~(7) provides a very accurate description of the number
equation near and at $T =0$, thus confirming Leggett's 
suggestion \cite{leggett-80b}.
However, it is well known that the same is not true near $T_c$, 
where the effects of temporal fluctuations are essential
to describe the BEC regime~\cite{sademelo-93}. 
The discussion of this interesting limit will be postponed 
to a future manuscript, and we will focus here on the low 
temperature properties, to be discussed next.

%%%%%%%%%%%%%% Momentum distribution %%%%%%%%%%%%%%%

{\it Momentum distribution:}
The momentum distribution $n_{\bf k}$ given by Eq.~(7), 
which at zero temperature reduces to
$
n_{\bf k} = (1 - \xi_{\bf k} / E_{\bf k}) / 2,
$
is plotted in Fig. 2 for the case of $p$-wave pairing 
symmetry as a function of ${\bf k} = (k_x , k_y)$, together 
with the contour plots. Notice that $n_{\bf k}$ becomes 
discontinuous when the chemical potential crosses zero, which 
coincides with the collapse of the two Dirac points to a single 
point ${\bf k} = 0$ and the appearence of a full gap in the quasiparticle 
excitation spectrum. This major rearrangement of the momentum
distribution has a dramatic effect in the atomic 
compressibility, which is discussed next.

%%%%%%%%%%%%%% Compressibility %%%%%%%%%%%%%%%

{\it Atomic Compressibility:}
The first derivative of the chemical potential with respect to the 
density $n = N / L^2$ becomes non-analytic at the critical value 
of the binding energy in the $p$-wave case. As a consequence, 
the isothermal atomic compressibility $\kappa$, defined by
\begin{equation}
\kappa = -{L^2\over N^2} {\partial^2\Omega \over \partial\mu^2}
= {1\over n^2} {\partial n \over \partial\mu},
\end{equation}
will develop a cusp when expressed in terms of $\tilde E_b$, 
its first derivative with respect to $\tilde E_b$ diverging 
at the critical point, as shown in Fig.~3. 
In the $s$-wave case, however, $\kappa$ is smooth for all
values of $\tilde E_b$~\cite{duncan-00}. 
This non-analytic behavior of the $p$-wave atomic 
compressibility, combined with the appearence of a full gap in 
the excitation spectrum, suggests the existence of a quantum critical 
point at $\tilde\mu = 0$.
\begin{figure}
\begin{center}
\includegraphics[width=6.5cm]{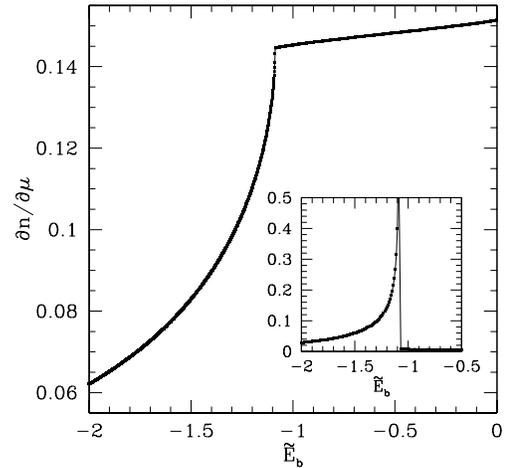}
\end{center}
\vspace{-5mm}
\caption{\small Plot of $\partial n / \partial \mu$ 
(in units of $k_F^2 / 4\pi\epsilon_F$) 
and its first derivative with respect to 
$\tilde E_b$ (inset)
as functions of $\tilde E_b$ in the case of spin polarized 
$p$-wave pairing and $k_0 = k_1 = 10 k_F$.}
\end{figure}
%%
%

%%%%%%%%%%%%%% Spin Susceptibility %%%%%%%%%%%%%%%%%
{\it Spin Susceptibility:} 
The phase transition discussed in the previous section also 
manifests itself in the spin susceptibility. 
The application of a small probe magnetic field 
$H_{\tilde z}$ along the same direction ($\tilde z$) 
of ${\bf h}$ generates the {\it spin} susceptibility response 
$
\chi_{\tilde z \tilde z} = (- 1/L^2)
(\partial^2 \Omega / \partial H_{\tilde z}^2) ,
$
which can be rewritten in the case of spin polarized atoms as
\begin{equation}
\chi_{\tilde z \tilde z} = 
- {1\over L^2} g_{\tilde z \tilde z}^2 \mu_B^2
{\partial^2 \Omega \over \partial \mu^2} = 
g_{\tilde z \tilde z}^2 \mu_B^2 
{\partial n \over \partial \mu}.
\end{equation}
Thus, the graph in Fig.~3 also represents a universal plot of 
$\chi_{\tilde z \tilde z}/g_{\tilde z \tilde z}^2 \mu_B^2$
as a function of ${\tilde E}_b$.

%%%%%%%%%%%%%% Superfluid Density %%%%%%%%%%%%%%%

{\it Superfluid Density:} 
We now turn our attention to the behavior of the low temperature
superfluid dentity tensor 
$
\rho_{ij}(T, \tilde E_b)
$
as the critical value of the binding energy $\tilde E_b$ is crossed.
This tensor is associated with phase twists of the superconductor 
order parameter \cite{feynman-72}, and can be obtained by taking 
$\phi_{\bf q} \to \phi_{\bf q} \exp(i\theta_{\bf q})$ 
and expanding the effective action $S_{\rm eff}$ in powers of 
$\theta_{\bf q}$ about the saddle point with $\theta_{\bf q}=0$.
The resulting difference in the action, 
$
\Delta S \equiv S_{\rm eff}(\theta_{\bf q}) - 
S_{\rm eff}(\theta_{\bf q}=0) ,
$
becomes
$
\Delta S(T) = -(L^2 / 2) \sum_{\bf q}
\theta_{\bf q} \theta_{-{\bf q}} q_i q_j
\rho_{ij}(T) ,
$
with the superfluid density tensor given by
\begin{equation}
\rho_{ij}(T) = {1\over 2 L^2} \sum_{\bf k} \left[
2 n_{\bf k} \partial_i \partial_j \xi_{\bf k} - 
Y_{\bf k} \partial_i \xi_{\bf k} \partial_j \xi_{\bf k}
\right] ,
\end{equation}
where $n_{\bf k}$ is the momentum distribution, 
$
Y_{\bf k} = (2T)^{-1} {\rm sech}^2(E_{\bf k}/2T)
$
is the Yoshida distribution, and $\partial_i$
denotes the partial derivative with respect to $k_i$.
Notice that $\rho_{xx} = \rho_{yy} \equiv \rho$, 
while $\rho_{xy} = \rho_{yx} = 0$. In addition, notice that
at $T=0$, $\rho_{ij}(0) = n/m$, such that 
$
\partial \rho_{ij} / \partial\mu = 
(1 / m)\partial n / \partial \mu
$
and 
$
\partial \rho_{ij} / \partial H_{\tilde z} = 
(1 / m)\partial n / \partial H_{\tilde z}.
$
Using our energy 
and momentum scales, we define the dimensionless quantity
$
\Delta\rho(T) \equiv m \rho(T) / n - 1 ,
$
which is shown in Fig.~4 as a function of temperature for different
values of the binding energy. The linear behavior of 
$\Delta\rho(T) / T^2$ for values of $\tilde E_b$ that correspond 
to $\tilde\mu > 0$ indicates a $T^3$ dependence of the superfluid 
density on temperature on the BCS side of the transition. This 
behavior is in fact confirmed by our analytical calculation of 
$\Delta\rho(T)$ at low temperatures and in the case of short range 
interactions ($k_0 \to \infty$). In the BCS limit, we found 
$\Delta\rho(T) \sim C T^3$, with the coefficient $C$ weakly 
dependent on $\tilde E_b$. This power-law behavior reflects the 
nodal (gapless) structure of the $p$-wave excitation spectrum.
In the BEC limit, we obtained
$\Delta\rho(T) \sim \exp(-|\tilde\mu| / T)$, the exponential 
behavior reflecting the appearance of a full gap to the addition of 
quasiparticles for $\tilde\mu < 0$. 
Fig.~4 also shows (inset) the zero temperature slope of 
$\Delta\rho(T) / T^2$ as a function of the binding energy $\tilde E_b$,
which is clearly discontinuous at the critical point 
$\tilde E_b = -1.087\epsilon_F$. These results further confirm  
the existence of a quantum phase transition along the BCS-to-BEC 
evolution as a function of interaction strength (binding energy)
in the case of $p$-wave spin polarized atoms.
\begin{figure}
\begin{center}
\includegraphics[width=6.5cm]{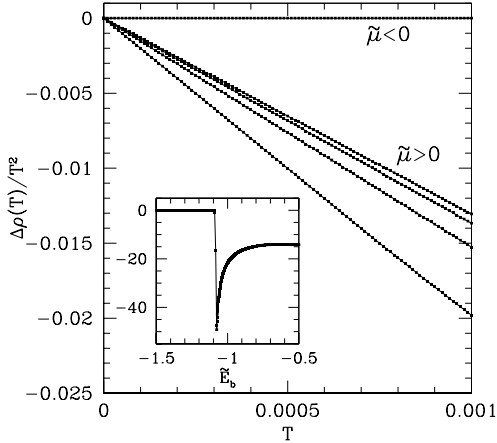}
\end{center}
\vspace{-5mm}
\caption{\small Plot of $\Delta\rho(T)/T^2$ 
(in units of $\epsilon_F^{-2}$) as a function of temperature
(in units of $\epsilon_F$) for various values of the binding energy 
$\tilde E_b$.
{\it Inset:} Zero-temperature slope of $\Delta\rho(T)/T^2$ 
(in units of $\epsilon_F^{-3}$) as a funcion of $\tilde E_b$.}
\end{figure}
%%
%

%%%%%%%%%%%%%% Summary %%%%%%%%%%%%%%%%%
{\it Summary:}
We proposed the existence of a quantum phase transition in
the BCS-to-BEC evolution of $p$-wave fully spin polarized Fermi 
gases as a function of the two-body bound state energy.
We have shown that, at a critical value of this binding energy, 
the momentum distribution undergoes a major rearrangement in 
${\bf k}$-space, which leads to a non-analytic behavior of 
the atomic compressibility and spin susceptibility of the gas. 
Furthermore, the low temperature superfluid density of the system
presents a dramatic change in behavior as the critical point 
is crossed, its temperature dependence going discontinuously 
from power-law on the BCS side of the transition to exponential 
on the BEC side of the transition. 

We conclude by suggesting that this phase
transition may be observable in traps of $^6{\rm Li}$ and 
$^{40}{\rm K}$ gases which exhibit $p$-wave Feshbach 
resonances~\cite{jin-03, zhang-04}. The occurrence of this 
phase transition may be investigated through the direct 
measurement of the atomic compressibility, spin susceptibility 
or superfluid density as functions of binding energy or 
magnetic field.

%%%%%%%%%%%%%% Acknowledgments %%%%%%%%%%%%%%%%%

{\it Acknowledgments:}
We would like to thank NSF (Grant No. DMR 0304380) for 
financial support, and Chandra Raman and Tony Leggett for 
references and discussions.

%%%%%%%%%%%%%% References %%%%%%%%%%%%%%%%%%%%%


\begin{references}

\bibitem{jin-04}
C. A. Regal, M. Greiner, and D. S. Jin,
Phys. Rev. Lett. {\bf 92}, 040403 (2004).

\bibitem{zwierlein-03}
M. W. Zwierlein {\it et al.}, 
Phys. Rev. Lett. {\bf 91}, 250401 (2003).

\bibitem{bourdel-04}
T. Bourdel {\it et al.}, 
cond-mat/0403091 (2004).

\bibitem{kinast-04}
J. Kinast {\it et al.}, 
Phys. Rev. Lett. {\bf 92}, 150402 (2004).

\bibitem{sademelo-93}
C. A. R. Sa de Melo, M. Randeria, and J. R. Engelbrecht, 
Phys. Rev. Lett. {\bf 71}, 3202 (1993).

\bibitem{engelbrecht-97}
J. R. Engelbrecht, M. Randeria, and C. A. R. Sa de Melo, 
Phys. Rev. B {\bf 55}, 15153 (1997).

\bibitem{eagles-69}
D. M. Eagles, 
Phys. Rev. {\bf 186}, 456 (1969).

\bibitem{leggett-80a} 
A. J. Leggett, in {\it Modern Trends in the Theory of Condensed
Matter}, edited by A. Peralski and R. Przystawa (Springer-Verlag,
Berlin, 1980).

\bibitem{nozieres-85}
P. Nozieres and S. Schmitt-Rink, J. Low Temp. Phys. {\bf 59},  
195 (1985).

\bibitem{holland-01}
M. Holland, S. J. J. M. F. Kokkelmans, 
M. L. Chiofalo, and R. Walser, 
Phys. Rev. Lett. {\bf 87}, 120406 (2001).

\bibitem{griffin-02}
Y. Ohashi and A. Griffin, 
Phys. Rev. Lett. {\bf 89}, 130402 (2002).

\bibitem{jin-03}
C. A. Regal, C. Ticknor, J. L. Bohn, and D. S. Jin,
Phys. Rev. Lett. {\bf 90}, 053201 (2003).

\bibitem{zhang-04}
J. Zhang {\it et al.}, 
quant-ph/0406085 (2004).

\bibitem{duncan-00}
R. D. Duncan and C. A. R. S\'a de Melo, 
Phys. Rev. B {\bf 62}, 9675 (2000).

\bibitem{randeria-89}
M. Randeria, J. Duan, and L. Shieh, 
Phys. Rev. Lett. {\bf 62}, 981 (1989).

\bibitem{leggett-80b}
A. J. Leggett, J. Phys. C (Paris) {\bf 41}, 7 (1980).

\bibitem{feynman-72}
R. P. Feynman, {\it Statistical Mechanics}
(W. A. Benjamin, Inc., Reading, MA, 1972), Chap. 10.

\end{references}
\end{document}